\documentclass[prc,twocolumn,aps,superscriptaddress,showpacs]{revtex4}
\usepackage{amssymb}
\usepackage{amsmath,bm}
\usepackage{graphicx}
\usepackage[normalem]{ulem}
\usepackage[dvips]{color}

\renewcommand\sout{\bgroup \color{red} \ULdepth=-.5ex \ULset}

\begin{document}

\title{Correlations between the nuclear breathing mode energy and properties
of asymmetric nuclear matter}
\author{Lie-Wen Chen}
\affiliation{Department of Physics, Shanghai Jiao Tong University, Shanghai 200240, China}
\affiliation{Center of Theoretical Nuclear Physics, National Laboratory of Heavy Ion
Accelerator, Lanzhou 730000, China}
\author{Jian-Zhong Gu}
\affiliation{China Institute of Atomic Energy, P. O. Box 275(10), Beijing 102413, China}
\date{\today}

\begin{abstract}
Based on microscopic Hartree-Fock + random phase approximation calculations
with Skyrme interactions, we study the correlations between the nuclear
breathing mode energy $E_{\mathrm{ISGMR}}$ and properties of asymmetric
nuclear matter with a recently developed analysis method. Our results
indicate that the $E_{\mathrm{ISGMR}}$ of $^{208}$Pb exhibits moderate
correlations with the density slope $L$ of the symmetry energy and the
isoscalar nucleon effective mass $m_{s,0}^{\ast }$ besides a strong
dependence on the incompressibility $K_{0}$ of symmetric nuclear matter.
Using the present empirical values of $L=60\pm 30$ MeV and $m_{s,0}^{\ast
}=(0.8\pm 0.1)m$, we obtain a theoretical uncertainty of about $\pm 16$ MeV
for the extraction of $K_{0}$ from the $E_{\mathrm{ISGMR}}$ of $^{208}$Pb.
Furthermore, we find the $E_{\mathrm{ISGMR}}$ difference between $^{100}$Sn
and $^{132}$Sn strongly correlates with $L$ and thus provides a potentially
useful probe of the symmetry energy.
\end{abstract}

\maketitle

\section{Introduction}

Determination of the equation of state (EOS) for isospin asymmetric nuclear
matter (ANM) is among fundamental questions in both nuclear physics and
astrophysics. Knowledge on the nuclear EOS is important for understanding
not only the structure of finite nuclei, the nuclear reaction dynamics, and
the liquid-gas phase transition in nuclear matter, but also many critical
issues such as properties of neutron stars and supernova explosion mechanism
in astrophysics~\cite{LiBA98,Dan02,Lat04,Ste05,Bar05,LCK08}. In the past
more than $30$ years, significant progress has been made in determining the
EOS of symmetric nuclear matter from subsaturation density to about $5$
times normal nuclear matter density $\rho _{0}$ by studying the nuclear
isoscalar giant monopole resonances (ISGMR)~\cite{You99}, collective flows~%
\cite{Dan02} and subthreshold kaon production~\cite{Aic85,Fuc06a} in
nucleus-nucleus collisions. On the other hand, the isospin dependent part of
the nuclear EOS, characterized essentially by the nuclear symmetry energy $%
E_{\text{\textrm{sym}}}({\rho })$, is still largely uncertain~\cite%
{Bar05,LCK08}. Lack of knowledge on the symmetry energy actually hinders us
to extract more accurately the EOS of symmetric nuclear matter. Therefore,
to explore and narrow down the uncertainties of both the theoretical methods
and the experimental data is of crucial importance for extracting more
stringently information on the nuclear EOS.

During the past more than 30 years, it has been established that the
nuclear ISGMR provides a good tool to probe the nuclear EOS around
the nuclear normal density. In particular, it is generally believed
that the incompressibility $K_{0}$ of symmetric nuclear matter can
be extracted from a self-consistent microscopic theoretical model
that successfully reproduces the experimental ISGMR energies as well
as the
ground state binding energies and charge radii of a variety of nuclei~\cite%
{Bla80}. Experimentally, thanks to new and improved experimental facilities
and techniques, the ISGMR centroid energy $E_{\mathrm{ISGMR}}$, i.e., the
so-called nuclear breathing mode energy, of $^{208}$Pb (a heavy,
doubly-magic nucleus with a well-developed monopole peak) has been measured
with a very high precision (less than $2\%$). Indeed, a value of $E_{\mathrm{%
ISGMR}}=14.17\pm 0.28$ MeV was extracted from the giant monopole resonance
in $^{208}$Pb based on an improved $\alpha $-scattering experiment~\cite%
{You99} (another value of $E_{\mathrm{ISGMR}}=13.96\pm 0.20$ MeV was
extracted in Ref.~\cite{You04}). The $E_{\mathrm{ISGMR}}$ of $^{208}$Pb has
been extensively used to constrain the $K_{0}$ parameter in the literature~%
\cite{You99,You04,Lui04,Ma02,Vre03,Col04,Shl06,LiT07,Gar07,Paa07,Col09,Pie10}%
. It is thus important to estimate and eventually narrow down the
theoretical uncertainty of extracting $K_{0}$ from the nuclear ISGMR.
Theoretically, in fact, it has been found that the uncertainty of the
density dependence of the symmetry energy has significantly influenced the
precise extraction of the $K_{0}$ parameter from ISGMR in $^{208}$Pb and it
also provides an explanation for the observed model dependence of the $K_{0}$
extraction from the ISGMR in $^{208}$Pb based on non-relativistic and
relativistic models~\cite{Pie02,Agr03,Vre03,Col04,Pie04}.

In the present work, we estimate the theoretical uncertainty when
one extracts the $K_{0}$ parameter from the nuclear ISGMR based on
microscopic Hartree-Fock (HF) + random phase approximation (RPA)
calculations with Skyrme interactions. In particular, we study the
correlations between the ISGMR centroid energy and properties of ANM
with a recently developed analysis method~\cite{Che10} in which
instead of varying directly the $9$ parameters in the Skyrme
interaction, we express them explicitly in terms of $9$ macroscopic
quantities that are either experimentally well constrained or
empirically well known. Then, by varying individually these
macroscopic quantities within their known ranges, we can examine
more transparently the correlation of the ISGMR centroid energy with
each individual macroscopic quantity and thus estimate the
theoretical uncertainty of the ISGMR centroid energy based on the
empirical uncertainties of the macroscopic quantities. Our results
indicate that the density slope $L$ of the symmetry energy and the
isoscalar nucleon effective mass $m_{s,0}^{\ast }$ can significantly
change the $E_{\mathrm{ISGMR}}$ of $^{208}$Pb and the present
uncertainties
of $L$ and $m_{s,0}^{\ast }$ can lead to a theoretical uncertainty of about $%
\pm 16$ MeV for the extraction of $K_{0}$. We further find the $E_{\mathrm{%
ISGMR}}$ difference between $^{100}$Sn and $^{132}$Sn displays a strong
correlation with $L$ and thus provides a potential probe of the symmetry
energy.

\section{Methods}

\label{Theory}

\subsection{Skyrme-Hartree-Fock approach and macroscopic properties of
asymmetric nuclear matter}

The EOS of isospin asymmetric nuclear matter, given by its binding energy
per nucleon, can be expanded to $2$nd-order in isospin asymmetry $\delta $
as
\begin{equation}
E(\rho ,\delta )=E_{0}(\rho )+E_{\mathrm{sym}}(\rho )\delta ^{2}+O(\delta
^{4}),  \label{EOSANM}
\end{equation}%
where $\rho =\rho _{n}+\rho _{p}$ is the baryon density with $\rho _{n}$ and
$\rho _{p}$ denoting the neutron and proton densities, respectively; $\delta
=(\rho _{n}-\rho _{p})/(\rho _{p}+\rho _{n})$ is the isospin asymmetry; $%
E_{0}(\rho )=E(\rho ,\delta =0)$ is the binding energy per nucleon in
symmetric nuclear matter, and the nuclear symmetry energy is expressed as
\begin{equation}
E_{\mathrm{sym}}(\rho )=\frac{1}{2!}\frac{\partial ^{2}E(\rho ,\delta )}{%
\partial \delta ^{2}}|_{\delta =0}.  \label{Esym}
\end{equation}%
Around $\rho _{0}$, the symmetry energy can be characterized by using the
value of $E_{\text{\textrm{sym}}}({\rho _{0}})$ and the density slope
parameter $L=3{\rho _{0}}\frac{\partial E_{\mathrm{sym}}(\rho )}{\partial
\rho }|_{\rho ={\rho _{0}}}$, i.e.,

\begin{equation}
E_{\mathrm{sym}}(\rho )=E_{\text{\textrm{sym}}}({\rho _{0}})+\frac{L}{3}(%
\frac{\rho -{\rho _{0}}}{{\rho _{0}}})+O((\frac{\rho -{\rho _{0}}}{{\rho _{0}%
}})^{2}).
\end{equation}

In the standard Skyrme Hartree-Fock approach, the nuclear effective
interaction is taken to have a zero-range, density- and momentum-dependent
form \cite{Cha97}, i.e.,
\begin{eqnarray}
V_{12}(\mathbf{R},\mathbf{r}) &=&t_{0}(1+x_{0}P_{\sigma })\delta (\mathbf{r})
\notag \\
&+&\frac{1}{6}t_{3}(1+x_{3}P_{\sigma })\rho ^{\sigma }(\mathbf{R})\delta (%
\mathbf{r})  \notag \\
&+&\frac{1}{2}t_{1}(1+x_{1}P_{\sigma })(\mathbf{K}^{^{\prime }2}\delta (%
\mathbf{r})+\delta (\mathbf{r})\mathbf{K}^{2})  \notag \\
&+&t_{2}(1+x_{2}P_{\sigma })\mathbf{K}^{^{\prime }}\cdot \delta (\mathbf{r})%
\mathbf{K}  \notag \\
&\mathbf{+}&iW_{0}(\mathbf{\sigma }_{1}+\mathbf{\sigma }_{2})\cdot \lbrack
\mathbf{K}^{^{\prime }}\times \delta (\mathbf{r})\mathbf{K]},  \label{V12Sky}
\end{eqnarray}%
with $\mathbf{r}=\mathbf{r}_{1}-\mathbf{r}_{2}$ and $\mathbf{R}=(\mathbf{r}%
_{1}+\mathbf{r}_{2})/2$. In the above expression, the relative momentum
operators $\mathbf{K}=(\mathbf{\nabla }_{1}-\mathbf{\nabla }_{2})/2i$ and $%
\mathbf{K}^{\prime }=-(\mathbf{\nabla }_{1}-\mathbf{\nabla }_{2})/2i$ act on
the wave function on the right and left, respectively. The quantities $%
P_{\sigma }$ and $\sigma _{i}$ denote, respectively, the spin exchange
operator and Pauli spin matrices. The $\sigma $, $t_{0}-t_{3}$, $x_{0}-x_{3}$
are the $9$ Skyrme interaction parameters which can be expressed
analytically in terms of $9$ macroscopic quantities $\rho _{0}$, $E_{0}(\rho
_{0})$, the incompressibility $K_{0}$, the isoscalar effective mass $%
m_{s,0}^{\ast }$, the isovector effective mass $m_{v,0}^{\ast }$, $E_{\text{%
\textrm{sym}}}({\rho _{0}})$, $L$, the gradient coefficient $G_{S}$, and the
symmetry-gradient coefficient $G_{V}$~\cite{Che10,Che11b}, i.e.,
\begin{eqnarray}
t{_{0}} &=&4\alpha /(3{\rho _{0}}) \\
x{_{0}} &=&3(y-1)E_{\text{\textrm{sym}}}^{\mathrm{loc}}({\rho _{0}})/\alpha
-1/2 \\
t{_{3}} &=&16\beta /\left[ {\rho _{0}}^{\gamma }(\gamma +1)\right] \\
x{_{3}} &=&-3y(\gamma +1)E_{\text{\textrm{sym}}}^{\mathrm{loc}}({\rho _{0}}%
)/(2\beta )-1/2 \\
t_{1} &=&20C/\left[ 9{\rho _{0}(}k_{\mathrm{F}}^{0})^{2}\right] +8G_{S}/3 \\
t_{2} &=&\frac{4(25C-18D)}{9{\rho _{0}(}k_{\mathrm{F}}^{0})^{2}}-\frac{%
8(G_{S}+2G_{V})}{3} \\
x_{1} &=&\left[ 12G_{V}-4G_{S}-\frac{6D}{{\rho _{0}(}k_{\mathrm{F}}^{0})^{2}}%
\right] /(3t_{1}) \\
x_{2} &=&\left[ 20G_{V}+4G_{S}-\frac{5(16C-18D)}{3{\rho _{0}(}k_{\mathrm{F}%
}^{0})^{2}}\right] /(3t_{2}) \\
\text{\ }\sigma &=&\gamma -1  \label{SkySigma}
\end{eqnarray}%
where $k_{\mathrm{F}}^{0}=(1.5\pi ^{2}{\rho _{0}})^{1/3}$, $E_{\text{\textrm{%
sym}}}^{\mathrm{loc}}({\rho _{0}})=E_{\text{\textrm{sym}}}({\rho _{0}})-E_{%
\text{\textrm{sym}}}^{\mathrm{kin}}({\rho _{0}})-D$, and the parameters $C$,
$D$, $\alpha $, $\beta $, $\gamma $, and $y$ are defined as \cite{Che09}
\begin{eqnarray}
C &=&\frac{m-m_{s,0}^{\ast }}{m_{s,0}^{\ast }}E_{\mathrm{kin}}^{0} \\
D &=&\frac{5}{9}E_{\mathrm{kin}}^{0}\left( 4\frac{m}{m_{s,0}^{\ast }}-3\frac{%
m}{m_{v,0}^{\ast }}-1\right) \\
\alpha &=&-\frac{4}{3}E_{\mathrm{kin}}^{0}-\frac{10}{3}C-\frac{2}{3}(E_{%
\mathrm{kin}}^{0}-3E_{0}(\rho _{0})-2C)  \notag \\
&&\times \frac{K_{0}+2E_{\mathrm{kin}}^{0}-10C}{K_{0}+9E_{0}(\rho _{0})-E_{%
\mathrm{kin}}^{0}-4C} \\
\beta &=&(\frac{E_{\mathrm{kin}}^{0}}{3}-E_{0}(\rho _{0})-\frac{2}{3}C)
\notag \\
&&\times \frac{K_{0}-9E_{0}(\rho _{0})+5E_{\mathrm{kin}}^{0}-16C}{%
K_{0}+9E_{0}(\rho _{0})-E_{\mathrm{kin}}^{0}-4C} \\
\gamma &=&\frac{K_{0}+2E_{\mathrm{kin}}^{0}-10C}{3E_{\mathrm{kin}%
}^{0}-9E_{0}(\rho _{0})-6C}. \\
y &=&\frac{L-3E_{\text{\textrm{sym}}}({\rho _{0}})+E_{\text{\textrm{\ sym}}%
}^{\mathrm{kin}}({\rho _{0}})-2D}{3(\gamma -1)E_{\text{ \textrm{sym}}}^{%
\mathrm{loc}}({\rho _{0}})}
\end{eqnarray}%
with $E_{\mathrm{kin}}^{0}=\frac{3\hbar ^{2}}{10m}\left( \frac{3\pi ^{2}}{2}%
\right) ^{2/3}\rho _{0}^{2/3}$ and $E_{\text{\textrm{sym}}}^{\mathrm{kin}}({%
\rho _{0}})=\frac{\hbar ^{2}}{6m}\left( \frac{3\pi ^{2}}{2}{\rho _{0}}%
\right) ^{2/3}$. In the above, the isoscalar effective mass $m_{s,0}^{\ast }$
is the nucleon effective mass in symmetric nuclear matter at its saturation
density ${\rho _{0}}$ while the isovector effective mass $m_{v,0}^{\ast }$
corresponds to the proton (neutron) effective mass in pure neutron (proton)
matter at baryon number density ${\rho _{0}}$ (See, e.g., Refs. \cite%
{Cha97,Che09}). Furthermore, $G_{S}$ and $G_{V}$ are respectively the
gradient and symmetry-gradient coefficients in the interaction part of the
binding energies for finite nuclei defined as
\begin{equation}
E_{\mathrm{grad}}=G_{S}(\nabla \rho )^{2}/(2{\rho )}-G_{V}\left[ \nabla
(\rho _{n}-\rho _{p})\right] ^{2}/(2{\rho )}.
\end{equation}

\subsection{HF + continuum-RPA calculations}

Since the energy of the giant monopole resonance is above the single
particle continuum threshold, a proper calculation should, in principle,
involve a complete treatment of the particle continuum. In the present work,
we study the ISGMR of nuclei by using a microscopic HF + continuum-RPA
calculations with Skyrme interactions \cite{Ham96}. The RPA response
function is solved in the coordinate space with the proton-neutron formalism
including simultaneously both the isoscalar and the isovector correlation.
In this way, we can take properly into account the coupling to the continuum
and the effect of neutron (proton) excess on the structure of the giant
resonances in nuclei near the neutron~(proton) drip lines~\cite{Ham96}.

The RPA strength distribution of ISGMR of nuclei
\begin{equation}
S(E_{x})=\sum_{n}|<n|Q|0>|^{2}\delta (E_{x}-E_{n})  \label{RPAStrength}
\end{equation}%
can be obtained by using the isoscalar monopole operator
\begin{equation}
Q^{\lambda =0,\tau =0}=\frac{1}{\sqrt{4\pi }}\sum_{i}r_{i}^{2}.
\label{ISGMROperator}
\end{equation}%
Furthermore, one can define the \textsl{k}-th energy moment of the
transition strength by
\begin{equation}
m_{k}=\int dE_{x}E_{x}^{k}S(E_{x}).  \label{RPAmoment}
\end{equation}%
The average energy of ISGMR can be defined by the ratio between the moments $%
m_{1}$ and $m_{0}$, i.e.,
\begin{equation}
E_{ave}=m_{1}/m_{0}.  \label{Eave}
\end{equation}%
In addition, the so-called scaling energy of ISGMR can be expressed as
\begin{equation}
E_{sca}=\sqrt{m_{3}/m_{1}},  \label{Esca}
\end{equation}%
while the ISGMR energy obtained from the constrained HF approach can be
written as
\begin{equation}
E_{con}=\sqrt{m_{1}/m_{-1}}.  \label{Econ}
\end{equation}%
The ISGMR energies defined by Eqs. (\ref{Eave})-(\ref{Econ}) will become
identical in the case of a sharp single peak exhausting $100\%$ of the sum
rule. In practice, it is found that both the experimental data and the
theoretical calculations show a large width of a few MeV even in the most
well-established ISGMR in $^{208}$Pb. However, it is interesting to note
that $E_{ave}$ and $E_{con}$ are rather close within a $0.1\sim 0.2$ MeV
difference even when the ISGMR peak has a large width although the scaling
energy $E_{sca}$ has a large uncertainty due to the high energy tail of
monopole strength, which is always the case in experimental data (and see
the theoretical results in the following). Furthermore, from the relation of
the energy moments $m_{k+1}m_{k-1}\geq m_{k}^{2}$, one can obtain $%
E_{sca}\geq E_{ave}\geq E_{con}$. Therefore, the average energy $E_{ave}$ is
usually defined as the ISGMR centroid energy and compared between the
experimental data and the theoretical calculations. It should be noted that
the situation of light nuclei may be quite different from that of medium and
heavy nuclei considered in the present work since the strength distribution
of ISGMR for light nuclei is usually very fragmented \cite{Lui01,Joh03,You09}%
. It was suggested in a recent work \cite{Fur10} that the fragmentation of
the strength distribution for the light nuclei might be explained by the
clustering effects which are not considered in the present work.

\section{Results}

\label{Result}

In the present work, as a reference for the correlation analyses performed
below with the standard Skyrme interactions, we use the MSL0 parameter set~%
\cite{Che10}, which is obtained by using the following empirical values for
the $9$ macroscopic quantities: $\rho _{0}=0.16$ fm$^{-3}$, $E_{0}(\rho
_{0})=-16$ MeV, $K_{0}=230$ MeV, $m_{s,0}^{\ast }=0.8m$, $m_{v,0}^{\ast
}=0.7m$, $E_{\text{\textrm{sym}}}({\rho _{0}})=30$ MeV, $L=60$ MeV, $G_{V}=5$
MeV$\cdot $fm$^{5}$, and $G_{S}=132$ MeV$\cdot $fm$^{5}$. And the spin-orbit
coupling constant $W_{0}=133.3$ MeV $\cdot $fm$^{5}$ is used to fit the
neutron $p_{1/2}-p_{3/2}$ splitting in $^{16}$O. It has been shown~\cite%
{Che10} that the MSL0 interaction can describe reasonably the binding
energies and charge rms radii for a number of closed-shell or
semi-closed-shell nuclei. We further find that the MSL0 parameter set
predicts a value of $1.06$ MeV for the splitting of the neutron $3p$ shell
in $^{208}$Pb, which reasonably describes the experimental value of $0.9$
MeV. It should be pointed out that the MSL0 is only used here as a reference
for the correlation analyses. Using other Skyrme interactions obtained from
fitting measured binding energies and charge rms radii of finite nuclei as
in usual Skyrme parametrization will not change our conclusion.

As numerical examples, in the present work, we choose the spherical
closed-shell doubly-magic nuclei $^{208}$Pb, $^{100}$Sn, and
$^{132}$Sn. Thus, we do not include the pairing interaction since it
has negligible effects on these spherical closed-shell doubly-magic
nuclei considered in this work \cite{Kha10}. In addition, the
two-body spin-orbit and the two-body Coulomb interactions are not
taken into account in the present continuum-RPA calculations
although the HF calculations include both of the interactions. As
pointed out in Ref. \cite{Sil06}, the net effect of the two
interactions in RPA decreases the centroid energy of ISGMR in
$^{208}$Pb by about $300$ keV. It should be stressed that, in the
present work, we do not intend to extract accurately the value of
the $K_{0}$ parameter by comparing the measured ISGMR centroid
energy with that from HF + continuum-RPA calculations, and our main
motivation is to explore the theoretical uncertainty for extracting
$K_{0}$. Meanwhile, we are mainly interested in the ISGMR centroid
energy difference between $^{100}$Sn and $^{132}$Sn rather than
their respective absolute value. Therefore, we do not expect that
the two-body spin-orbit and the two-body Coulomb interactions in RPA
will significantly change our conclusion and further work is needed
to see how exactly the two interactions in continuum-RPA
calculations will affect
our results. Furthermore, in the following calculations, the sum rules $%
m_{k} $ are obtained by integrating the RPA strength from excitation energy $%
E_{x}=5$ MeV to $E_{x}=35$ MeV in Eq. (\ref{RPAmoment}).

\subsection{Isospin scalar giant monopole resonance in $^{208}$Pb}

\begin{figure}[tbp]
\includegraphics[scale=0.77]{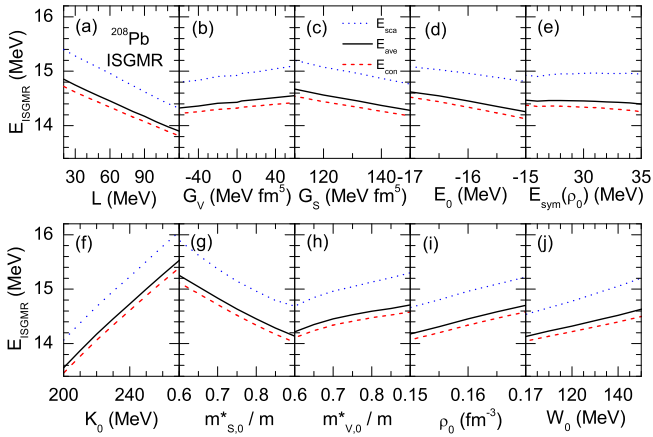}
\caption{(Color online) The ISGMR energies of $^{208}$Pb obtained from SHF +
continuum-RPA calculations with MSL0 by varying individually $L$ (a), $G_{V}$
(b), $G_{S}$ (c), $E_{0}(\protect\rho _{0})$ (d), $E_{\text{\textrm{sym}}}(%
\protect\rho _{0})$ (e), $K_{0}$ (f), $m_{s,0}^{\ast }$ (g), $m_{v,0}^{\ast
} $ (h), $\protect\rho _{0}$ (i), and $W_{0}$ (j). The three lines from
upper to lower in each panel correspond to $E_{sca}$, $E_{ave}$, and $%
E_{con} $, respectively.}
\label{XEaPb208}
\end{figure}
Shown in Fig. \ref{XEaPb208} are the ISGMR energies, i.e., $E_{sca}$, $%
E_{ave}$, and $E_{con}$ of $^{208}$Pb obtained from SHF + continuum-RPA
calculations with MSL0 by varying individually $L$, $G_{V}$, $G_{S}$, $%
E_{0}(\rho _{0})$, $E_{\text{\textrm{sym}}}(\rho _{0})$, $K_{0}$, $%
m_{s,0}^{\ast }$, $m_{v,0}^{\ast }$, $\rho _{0}$, and $W_{0}$, namely,
varying one quantity at a time while keeping all the others at their default
values in MSL0. Firstly, one can see clearly the ordering of $E_{sca}\geq
E_{ave}\geq E_{con}$ as expected. In particular, for the default parameters
in MSL0, we obtain $E_{sca}=14.962$ MeV, $E_{ave}=14.453$ MeV, and $%
E_{con}=14.338$ MeV. We note that the centroid energy of ISGMR $%
E_{ave}=14.453$ MeV is essentially in agreement with the measured value of $%
14.17\pm 0.28$ MeV for the ISGMR in $^{208}$Pb \cite{You99} (a more recent
experimental value of $13.96\pm 0.20$ MeV was extracted in Ref. \cite{You04}%
). The agreement will become much better if the two-body spin-orbit
and the two-body Coulomb interactions are taken into account in the
continuum-RPA calculations since the net effect of the two
interactions in RPA reduces the centroid energy of ISGMR in
$^{208}$Pb by about $300$ keV \cite{Sil06}. These features imply
that the default Skyrme parameter set MSL0 can give a good
description for the ISGMR in $^{208}$Pb. Furthermore, one can see
from Fig. \ref{XEaPb208} that, within the uncertain ranges
considered here for the macroscopic quantities, the ISGMR energies
display a very strong positive correlation with $K_{0}$ as expected.
On the other hand, however, the ISGMR energies also exhibit moderate
negative correlations with both $L$ and $m_{s,0}^{\ast }$ while weak
dependence on the other macroscopic quantities. These results
indicate that the uncertainties of $L$ and $m_{s,0}^{\ast }$ may
significantly influence the extraction of $K_{0}$ by comparing the
theoretical value of the ISGMR energies of $^{208}$Pb from SHF + RPA
calculations with the experimental measurements.
\begin{figure}[tbp]
\includegraphics[scale=1.0]{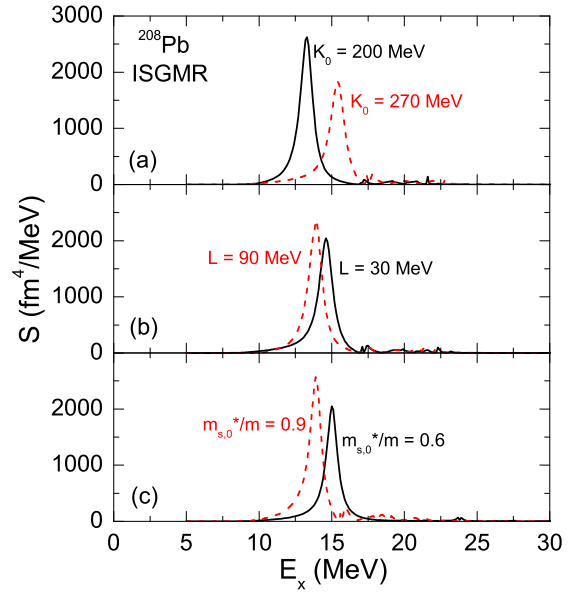}
\caption{(Color online) SHF + continuum-RPA response functions of $^{208}$Pb
with Skyrme interaction MSL0 by varying individually $K_{0}$ (a), $L$ (b),
and $m_{s,0}^{\ast }$ (c).}
\label{SRPAPb208}
\end{figure}

As for the correlation analysis method in Fig. \ref{XEaPb208}, we
would like to stress that, when the macroscopic quantities (except
$E_{0}(\rho _{0})$ and $\rho _{0}$) change individually from their
base values in MSL0 within the empirical uncertain ranges considered
here, the values of the binding energy or the charge rms radii of
finite nuclei will vary by only about $2\%$ at most (See, e.g.,
Figs. 4 and 5 in Ref. \cite{Che10}). Therefore, the original
agreement of MSL0 with the experimental data of binding energies or
charge radii of finite nuclei will essentially still hold with the
individual change of the macroscopic quantities. In this way,
changing the macroscopic quantities individually in the present
correlation analysis approach is equivalent to constructing a number
of different Skyrme interaction sets which can give reasonable
description on the ground state binding energy and charge rms radii
of finite nuclei. The key point and the most important advantage of
the present analysis approach is that in the present correlation
analysis, one knows exactly what is the difference among different
Skyrme interaction sets constructed by using different macroscopic
quantities. Furthermore, it should be mentioned that the centroid
energy of ISGMR in heavy nuclei are not sensitive to the values of
$E_{0}(\rho _{0})$ and $\rho _{0}$ as shown in Fig. \ref{XEaPb208},
and thus in principle we can adjust $E_{0}(\rho _{0})$ and $\rho
_{0}$ to give better description for the ground state binding energy
and charge rms radii of finite nuclei without changing significantly
the results of ISGMR and thus the conclusions in the present work.

In order to see the dependence of the detailed structure of ISGMR in $^{208}$%
Pb\ on the values of $K_{0}$,\ $L$ and $m_{s,0}^{\ast }$, we show in Fig. %
\ref{SRPAPb208} the SHF + continuum-RPA response functions of $^{208}$Pb
with MSL0 by varying individually $K_{0}$, $L$, and $m_{s,0}^{\ast }$, i.e.,
$K_{0}=200$ and $270$ MeV, $L=30$ and $90$ MeV, and $m_{s,0}^{\ast }=0.6m$
and $0.9m$. As can be seen in Fig. \ref{SRPAPb208}, the RPA result displays
a single collective peak in each case, consistent with the experimental data
\cite{You04,Sag07b}. Furthermore, it is seen that varying the value of $%
K_{0} $ from $200$ MeV to $270$ MeV strongly shifts the single collective
peak from about $13.3$ MeV to $15.4$ MeV while varying the value of $L$ ($%
m_{s,0}^{\ast }$) from $30$ MeV ($0.6m$) to $90$ MeV ($0.9m$) shifts the
single collective peak from about $14.6$ ($15.0$) MeV to $13.9$ ($13.9$)
MeV. These results are consistent with the results shown in Fig. \ref%
{XEaPb208}. In addition, the calculated width with MSL0 by varying
individually $K_{0}$, $L$, and $m_{s,0}^{\ast }$ shows roughly the same
value as that of experimental data \cite{You04,Sag07b}. This agreement
implies that the width of ISGMR is essentially determined by the Landau
damping and the coupling to the continuum, which are properly taken into
account in the present calculations.

The ISGMR energy $E_{\mathrm{ISGMR}}$ is conventionally related to a finite
nucleus incompressibility $K_{A}(N,Z)$ for a nucleus with $N$ neutrons and $Z
$ protons ($A=N+Z$) by the following definition%
\begin{equation}
E_{\mathrm{ISGMR}}=\sqrt{\frac{\hbar ^{2}K_{A}(N,Z)}{m\left\langle
r^{2}\right\rangle }},  \label{EGMRKa}
\end{equation}%
where $m$ is the nucleon mass and $\left\langle r^{2}\right\rangle $ is the
mean square mass radius of the nucleus in the ground state. Similarly to the
semi-empirical mass formula, the finite nucleus incompressibility $K_{A}(N,Z)
$ is usually expanded as \cite{Bla80}
\begin{eqnarray}
K_{A}(N,Z) &=&K_{0}+K_{\mathrm{surf}}A^{-1/3}+K_{\mathrm{curv}}A^{-2/3}
\notag \\
&&+(K_{\tau }+K_{\mathrm{ss}}A^{-1/3})\left( \frac{N-Z}{A}\right) ^{2}
\notag \\
&&+K_{\mathrm{Coul}}\frac{Z^{2}}{A^{4/3}}+\cdot \cdot \cdot .  \label{KA1}
\end{eqnarray}%
Neglecting the curvature term $K_{\mathrm{curv}}$, the surface symmetry term
$K_{\mathrm{ss}}$ and the other higher-order terms of the finite nucleus
incompressibility $K_{A}(N,Z)$ in Eq. (\ref{KA1}), one can express $%
K_{A}(N,Z)$ as
\begin{eqnarray}
K_{A}(N,Z) &=&K_{0}+K_{\mathrm{surf}}A^{-1/3}+K_{\tau }\left( \frac{N-Z}{A}%
\right) ^{2}  \notag \\
&&+K_{\mathrm{Coul}}\frac{Z^{2}}{A^{4/3}},  \label{KA2}
\end{eqnarray}%
where $K_{0}$, $K_{\mathrm{surf}}$, $K_{\tau }$, and
$K_{\mathrm{coul}}$ represent the volume, surface, symmetry, and
Coulomb terms, respectively. The $K_{\tau }$ parameter is usually
thought to be equivalent to the isospin dependent part, i.e., the
$K_{\mathrm{sat,2}}$ parameter, of the isobaric incompressibility
coefficient of ANM (incompressibility evaluated at the saturation
density of ANM) defined as%
\begin{equation}
K_{\mathrm{sat}}(\delta )=K_{0}+K_{\mathrm{sat,2}}\delta ^{2}+O(\delta ^{4}).
\label{Ksat}
\end{equation}%
We would like to point out that the $K_{\mathrm{sat,2}}$ parameter is a
theoretically well-defined physical property of ANM \cite{Pie09,Che09} while
the value of the $K_{\tau }$ parameter may depend on the details of the
truncation scheme in Eq. (\ref{KA1}) \cite{Bla81,Sha88,Pea91,Shl93,Pea10}.
Here, we assume $K_{\mathrm{sat,2}}$ has similar influences on $K_{A}(N,Z)$
as $K_{\tau }$ and thus $K_{\mathrm{sat,2}}$ will affect the $E_{\mathrm{%
ISGMR}}$ through Eq. (\ref{EGMRKa}), and then we can analyze the $L$ and $%
m_{s,0}^{\ast }$ dependences of $E_{\mathrm{ISGMR}}$ from the correlations
of $K_{\mathrm{sat,2}}$ parameter with $L$ and $m_{s,0}^{\ast }$.
\begin{figure}[tbp]
\includegraphics[scale=0.9]{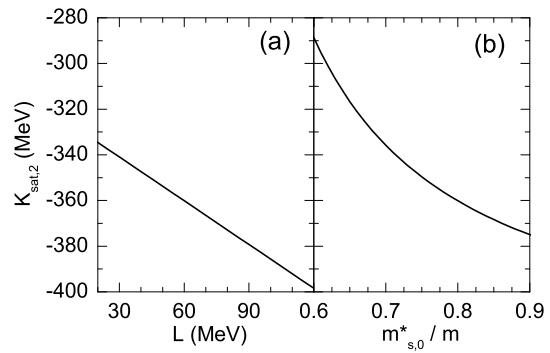}
\caption{The $K_{\mathrm{sat,2}}$ parameter obtained from SHF with MSL0 by
varying individually $L$ (a) and $m_{s,0}^{\ast }$ (b).}
\label{XKsat2}
\end{figure}

The effects of the density dependence of the symmetry energy on the ISGMR
centroid energy $E_{ave}$ of $^{208}$Pb has been extensively investigated in
the literature~\cite{Pie02,Vre03,Agr03,Pie04,Col04}. It was firstly proposed
by Piekarewicz~\cite{Pie02} that the different symmetry energies used in the
non-relativistic models and the relativistic models may be responsible for
the puzzle that the former predicted an incompressibility in the range of $%
K_{0}=210-230$ MeV while the latter predicted a significantly larger value
of $K_{0}\approx 270$ MeV from the analysis of the ISGMR centroid energy. It
is seen from Fig. \ref{XEaPb208} that a larger $L$ value (as in usual
relativistic models) leads to a smaller $E_{ave}$ value and thus a larger $%
K_{0}$ value is necessary to counteract the deceasing of $E_{ave}$ due to a
larger $L$ value. Furthermore, Fig. \ref{XEaPb208} shows that $E_{ave}$
displays a very weak dependence on $E_{\text{\textrm{sym}}}(\rho _{0})$,
which is in contrast to the results in Ref.~\cite{Col04} where $E_{ave}$ is
shown to be sensitive to $E_{\text{\textrm{sym}}}(\rho _{0})$. This is due
to the fact that a constrain on the value of $E_{\text{\textrm{sym}}}({\rho
=0.1}$ {fm}$^{{-3}})$ was imposed in Ref.~\cite{Col04}, which leads to a
strong linear correlation between $E_{\text{\textrm{sym}}}({\rho _{0}})$ and
$L$ as shown recently in Ref.~\cite{Che11a}.

The symmetry energy dependence of the ISGMR centroid energy of $^{208}$Pb
can be understood from the fact that the ISGMR in $^{208}$Pb does not
constrain the compression modulus of symmetric nuclear matter but rather the
one of neutron-rich matter, i.e., the isobaric incompressibility coefficient
in Eq. (\ref{Ksat}). From Eq. (\ref{Ksat}) it is clear that the ISGMR in $%
^{208}$Pb (with an isospin asymmetry of $\delta =0.21$) should be sensitive
to a linear combination of $K_{0}$ and $K_{\mathrm{sat,2}}$. The $K_{\mathrm{%
sat,2}}$ parameter is completely determined by the slope and curvature of
the symmetry energy at saturation density as well as the third derivative of
the EOS of symmetric nuclear matter (see, e.g., Ref. \cite{Che09}). Fig. \ref%
{XKsat2} shows the $K_{\mathrm{sat,2}}$ parameter from SHF with MSL0 by
varying individually $L$ and $m_{s,0}^{\ast }$. As can be seen in Fig. \ref%
{XKsat2}, the $K_{\mathrm{sat,2}}$ parameter decreases with both $L$ and $%
m_{s,0}^{\ast }$, and thus $K_{A}(N,Z)$ for $^{208}$Pb will decrease
correspondingly if the $K_{\mathrm{sat,2}}$ parameter has similar effects on
$K_{A}(N,Z)$ as the $K_{\tau }$ parameter and the $K_{\mathrm{ss}}$ term as
well as the other higher-order terms in Eq. (\ref{KA1}) are not important
for $K_{A}(N,Z)$. These results provide an explanation on the behavior that
the ISGMR energies decrease with $L$ and $m_{s,0}^{\ast }$ observed in Fig. %
\ref{XEaPb208}.
\begin{figure}[tbp]
\includegraphics[scale=1.0]{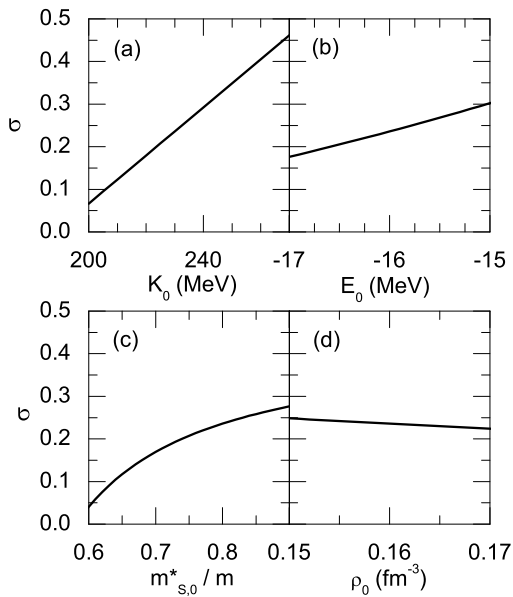}
\caption{The $\protect\sigma $ parameter obtained from SHF with MSL0 by
varying individually $K_{0}$ (a), $E_{0}(\protect\rho _{0})$ (b), $%
m_{s,0}^{\ast }$ (c), and $\protect\rho _{0}$ (d).}
\label{XSigmaPara}
\end{figure}

To understand more clearly why the ISGMR energies decrease with $%
m_{s,0}^{\ast }$ observed in Fig. \ref{XEaPb208}, it is useful to note the
fact that, with the standard Skyrme interaction, the $K_{0}$ and $%
m_{s,0}^{\ast }$ cannot be chosen independently if the Skyrme interaction
parameter $\sigma $ in Eq. (\ref{V12Sky}), $E_{0}(\rho _{0})$ and $\rho _{0}$
are fixed \cite{Boh79}. It should be stressed here that, instead of assuming
a fixed value of $\sigma $ as in the usual parametrization and correlation
analysis \cite{Cha97,Col04}, in the present work, the $\sigma $ parameter is
determined by four macroscopic quantities, i.e., $K_{0}$, $E_{0}(\rho _{0})$%
, $m_{s,0}^{\ast }$ and $\rho _{0}$ as shown in Eq. (\ref{SkySigma}), and
thus $K_{0}$ and $m_{s,0}^{\ast }$ can be chosen independently. Neglecting
the isospin dependence (assuming $N\approx Z)$, the nuclear breathing mode
energy for medium and heavy nuclei can be approximated by \cite{Boh79}
\begin{equation}
E_{\mathrm{ISGMR}}\approx \sqrt{\frac{\hbar ^{2}(K_{0}-63\sigma )}{%
m\left\langle r^{2}\right\rangle }}\text{ (}K_{0}\text{ in MeV).}
\label{EGMRBohigas}
\end{equation}%
Eq. (\ref{EGMRBohigas}) implies that\ the nuclear breathing mode energy can
be closely related to both $K_{0}$ and $m_{s,0}^{\ast }$ if the parameter $%
\sigma $ is free and the values of $E_{0}(\rho _{0})$ and $\rho _{0}$ are
fixed. In Fig. \ref{XSigmaPara}, we show the $\sigma $ parameter obtained
from SHF with MSL0 by varying individually $K_{0}$, $E_{0}(\rho _{0})$, $%
m_{s,0}^{\ast }$, and $\rho _{0}$. One can see clearly that the $\sigma $
parameter indeed exhibits a strong correlation with $K_{0}$ as expected.
However, it also displays a moderate dependence on $m_{s,0}^{\ast }$, a
small dependence on $E_{0}(\rho _{0})$, and a very weak correlation with $%
\rho _{0}$. As can be seen in Fig. \ref{XSigmaPara}, the $\sigma $ parameter
increases with $m_{s,0}^{\ast }$, leading to smaller ISGMR energies
according to Eq. (\ref{EGMRBohigas}), which is consistent with the results
shown in Fig. \ref{XEaPb208}. In addition, the fact that $K_{\mathrm{sat,2}}$
parameter decreases with $m_{s,0}^{\ast }$ observed in Fig. \ref{XKsat2}
will also be partially responsible for the behavior of ISGMR energies
decreasing with $m_{s,0}^{\ast }$ as seen in Fig. \ref{XEaPb208} since a
smaller $K_{\mathrm{sat,2}}$ value will lead to a smaller $E_{\mathrm{ISGMR}%
} $ as discussed previously.

The above results indicate that the ISGMR centroid energy of $^{208}$Pb
exhibits moderate correlations with both $L$ and $m_{s,0}^{\ast }$ besides a
strong dependence on $K_{0}$. The accurate knowledge on $L$ and $%
m_{s,0}^{\ast }$ is thus important for a precise determination of the $K_{0}$
parameter from the ISGMR centroid energy of $^{208}$Pb. In recent years,
significant progress has been made in determining $L$ and its value is
essentially consistent with $L=60\pm 30$ MeV depending on the observables
and methods used in the studies~\cite%
{Mye96,Che05a,She07,Kli07,Tri08,Tsa09,Dan09,Cen09,Car10,XuC10,Liu10,Che11a}.
Using $L=60\pm 30$ MeV, we can estimate an uncertainty of about $\pm 0.281$
MeV for the ISGMR centroid energy in $^{208}$Pb from Fig. \ref{XEaPb208}. On
the other hand, for the isoscalar effective mass, the empirical value of $%
m_{s,0}^{\ast }=(0.8\pm 0.1)m$ has been obtained from the analysis of both
isoscalar quadrupole giant resonances data in doubly closed-shell nuclei and
single-particle spectra \cite{Liu76,Boh79,Far97,Rei99,Les06}. From Fig. \ref%
{XEaPb208}, we can obtain an uncertainty of about $\pm 0.382$ MeV for the
ISGMR centroid energy in $^{208}$Pb using the empirical value of $%
m_{s,0}^{\ast }=(0.8\pm 0.1)m$. Assuming the two uncertainties due to the
present uncertainties of $L$ and $m_{s,0}^{\ast }$ on the ISGMR centroid
energy in $^{208}$Pb are independent, we thus can add them quadratically to
obtain an uncertainty of about $\pm 0.474$ MeV for the ISGMR centroid energy
in $^{208}$Pb. Then, using the approximate relation $(\delta K_{0}/K_{0})=2$(%
$\delta E_{\mathrm{ISGMR}}$/$E_{\mathrm{ISGMR}}$) from Eq. (\ref{EGMRKa}),
we can obtain an uncertainty of $\pm 7\%$ for $K_{0}$ with $E_{\mathrm{ISGMR}%
}\approx 14$ MeV, namely, about $\pm 16$ MeV for $K_{0}=230$ MeV.

Furthermore, including other uncertainties due to $G_{V}$, $G_{S}$, $%
E_{0}(\rho _{0})$, $E_{\text{\textrm{sym}}}(\rho _{0})$, $m_{v,0}^{\ast }$, $%
\rho _{0}$ and $W_{0}$ with empirical values of $G_{V}=0\pm 40$ MeV, $%
G_{S}=130\pm 10$ MeV, $E_{0}(\rho _{0})=-16\pm 1$ MeV, $E_{\text{\textrm{sym}%
}}(\rho _{0})=30\pm 5$ MeV, $m_{v,0}^{\ast }=(0.7\pm 0.1)m$, $\rho
_{0}=0.16\pm 0.01$ fm$^{-3}$ and $W_{0}=130\pm 20$ MeV, and assuming all the
uncertainties are independent, we can obtain from Fig. \ref{XEaPb208} a
total uncertainty of about $\pm 0.647$ MeV for the ISGMR centroid energy in $%
^{208}$Pb, which gives an uncertainty of about $\pm 9\%$ for $K_{0}$,
namely, about $\pm 21$ MeV for $K_{0}=230$ MeV.

\subsection{Isospin scalar giant monopole resonances in $^{100}$Sn and $%
^{132}$Sn}

\begin{figure}[tbp]
\includegraphics[scale=0.77]{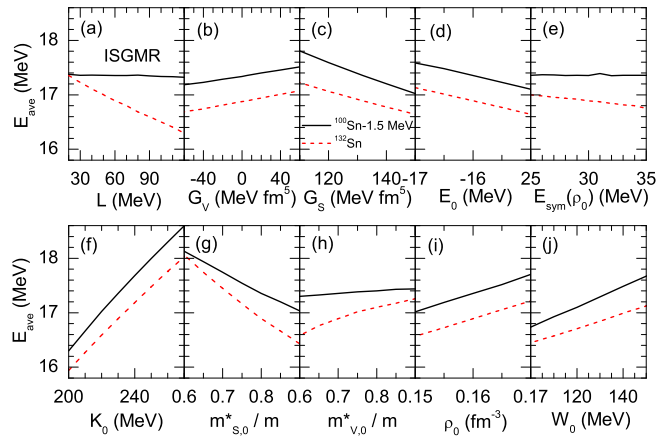}
\caption{(Color online) Same as Fig. \protect\ref{XEaPb208} but for the
ISGMR centroid energy $E_{ave}$ of $^{100}$Sn and $^{132}$Sn. The results of
$^{100}$Sn shift down by $1.5$ MeV for a more clear comparison with those of
$^{132}$Sn.}
\label{XEcSn100132}
\end{figure}

To see the isotopic dependence of the ISGMR centroid energy, we study here
the spherical closed-shell doubly-magic nuclei $^{100}$Sn and $^{132}$Sn.
Shown in Fig. \ref{XEcSn100132} are the ISGMR centroid energy $E_{ave}$ of $%
^{100}$Sn and $^{132}$Sn obtained from SHF + RPA calculations with MSL0 by
varying individually $L$, $G_{V}$, $G_{S}$, $E_{0}(\rho _{0})$, $E_{\text{%
\textrm{sym}}}(\rho _{0})$, $K_{0}$, $m_{s,0}^{\ast }$, $m_{v,0}^{\ast }$, $%
\rho _{0}$, and $W_{0}$. One can see that the results for neutron-rich
nucleus $^{132}$Sn are quite similar to those for $^{208}$Pb as shown in
Fig. \ref{XEaPb208}. On the other hand, for the symmetric nucleus $^{100}$%
Sn, it is interesting to see that the dependence of $E_{ave}$ on the isospin
relevant macroscopic quantities, namely, $L$, $G_{V}$, $E_{\text{\textrm{sym}%
}}(\rho _{0})$, $m_{v,0}^{\ast }$ is very weak. We have also checked the
case of the stable nucleus $^{90}$Zr for the correlation analysis as in Fig. %
\ref{XEcSn100132}, and we find the results are very similar to the case of $%
^{100}$Sn, namely, displaying a much weak correlation with the $L$
parameter while a stronger correlation with the $G_{S}$ parameter
compared with the case of $^{208}$Pb. This may be understandable
from the fact that the $^{90}$Zr has a smaller isospin asymmetry,
i.e., $(N-Z)/A=0.11$ compared with $^{208}$Pb where we have
($N-Z)/A=0.21$. In addition, the surface coefficient $G_{S}$ may
become more important for lighter nuclei as expected, leading to a
stronger correlation with the $G_{S}$ parameter. From these results,
it seems that the ISGMR of a heavier and more symmetric nucleus,
where the symmetry energy effects will be reduced significantly, may
be more suitable for extracting the $K_{0}$ parameter. In addition,
the different $E_{ave}$-$m_{s,0}^{\ast }$ correlations between
$^{100}$Sn and $^{132}$Sn observed in Fig. \ref{XEcSn100132} can be
understood from the fact that $K_{\mathrm{sat,2}}$ parameter
decreases with $m_{s,0}^{\ast }$ as shown in Fig. \ref{XKsat2},
leading additional decrement of $E_{ave}$ with $m_{s,0}^{\ast }$ for
the neutron-rich nucleus $^{132}$Sn.

\begin{figure}[tbp]
\includegraphics[scale=0.77]{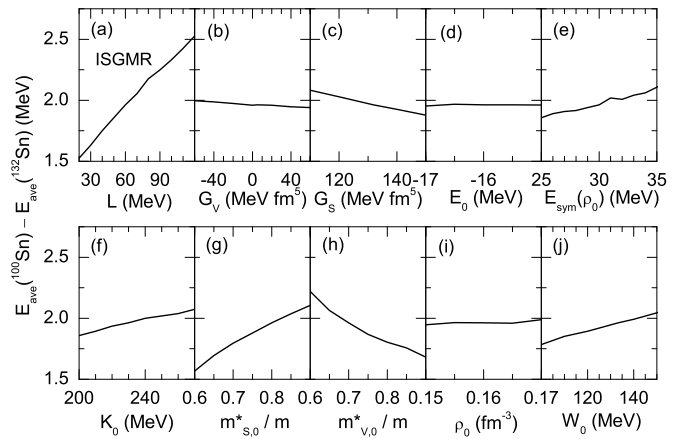}
\caption{Same as Fig. \protect\ref{XEaPb208} but for the ISGMR
centroid energy difference between $^{100}$Sn and $^{132}$Sn.}
\label{dEcSn100132}
\end{figure}
It is instructive to see the ISGMR centroid energy difference between $%
^{100} $Sn and $^{132}$Sn, which is shown in Fig. \ref{dEcSn100132} with
MSL0 by varying individually $L$, $G_{V}$, $G_{S}$, $E_{0}(\rho _{0})$, $E_{%
\text{\textrm{sym}}}(\rho _{0})$, $K_{0}$, $m_{s,0}^{\ast }$, $m_{v,0}^{\ast
}$, $\rho _{0}$, and $W_{0}$. It is very interesting to see from Fig. \ref%
{dEcSn100132} that, within the uncertain ranges considered here for the
macroscopic quantities, the ISGMR centroid energy difference displays a very
strong correlation with $L$. However, on the other hand, the ISGMR centroid
energy difference exhibits only moderate correlations with $m_{s,0}^{\ast }$
and $m_{v,0}^{\ast }$ while weak dependence on the other macroscopic
quantities. These features imply that the ISGMR centroid energy difference
between $^{100}$Sn and $^{132}$Sn provides a potential probe of the $L$
parameter. Furthermore, it is seen that the ISGMR centroid energy difference
displays opposite correlation with $m_{s,0}^{\ast }$ and $m_{v,0}^{\ast }$,
namely, increases with $m_{s,0}^{\ast }$ while decreases with $m_{v,0}^{\ast
}$. Recently, a constraint of $m_{s,0}^{\ast }-m_{v,0}^{\ast }=(0.126\pm
0.051)m$ has been extracted from global nucleon optical potentials
constrained by world data on nucleon-nucleus and (p, n) charge-exchange
reactions~\cite{XuC10}. Imposing the constraint $m_{s,0}^{\ast
}-m_{v,0}^{\ast }=(0.126\pm 0.051)m$, we can expect from Fig. \ref%
{dEcSn100132} that the correlation of the ISGMR centroid energy
difference with $m_{s,0}^{\ast }$ and $m_{v,0}^{\ast }$ will become
significantly weak, making the ISGMR centroid energy difference
really a good probe of the $L$ parameter. Our results indicate that
a precise determination of the ISGMR centroid energy difference
between $^{100}$Sn and $^{132}$Sn will be potentially useful to
constraint accurately the symmetry energy, especially the $L$
parameter. This provides strong motivation for measuring the ISGMR
strength in unstable nuclei, which can be investigated at the
new/planning rare isotope beam facilities at CSR/HIRFL and
BRIF-II/CIAE in China, RIBF/RIKEN in Japan, SPIRAL2/GANIL in France,
FAIR/GSI in Germany, and FRIB/NSCL in USA.

\section{Summary}

\label{Summary}

The isoscalar giant monopole resonances of finite nuclei have been
investigated based on microscopic Hartree-Fock + random phase approximation
calculations with Skyrme interactions. In particular, we have studied the
correlations between the ISGMR centroid energy, i.e., the so-called nuclear
breathing mode energy, and properties of asymmetric nuclear matter within a
recently developed correlation analysis method. Our results indicate that
the ISGMR centroid energy of $^{208}$Pb displays a very strong correlation
with $K_{0}$ as expected. On the other hand, however, the ISGMR centroid
energy also exhibits moderate correlation with both $L$ and $m_{s,0}^{\ast }$
while weak dependence on the other macroscopic quantities. Using the present
empirical values of $L=60\pm 30$ MeV and $m_{s,0}^{\ast }=(0.8\pm 0.1)m$, we
have obtained an uncertainty of about $0.474$ MeV for the ISGMR centroid
energy in $^{208}$Pb, leading to a theoretical uncertainty of about $\pm 16$
MeV for the extraction of $K_{0}$ from the $E_{\mathrm{ISGMR}}$ of $^{208}$%
Pb. Including additionally other uncertainties due to $G_{V}$, $G_{S}$, $%
E_{0}(\rho _{0})$, $E_{\text{\textrm{sym}}}(\rho _{0})$, $m_{v,0}^{\ast }$, $%
\rho _{0}$ and $W_{0}$ with empirical values of $G_{V}=0\pm 40$ MeV, $%
G_{S}=130\pm 10$ MeV, $E_{0}(\rho _{0})=-16\pm 1$ MeV, $E_{\text{\textrm{sym}%
}}(\rho _{0})=30\pm 5$ MeV, $m_{v,0}^{\ast }=(0.7\pm 0.1)m$, $\rho
_{0}=0.16\pm 0.01$ fm$^{-3}$ and $W_{0}=130\pm 20$ MeV, we have estimated a
total uncertainty of about $\pm 21$ MeV for the extraction of $K_{0}$ by
assuming all the uncertainties are independent. These results show that the
accurate knowledge on $L$ and $m_{s,0}^{\ast }$ is important for a precise
determination of the $K_{0}$ parameter by comparing the measured ISGMR
centroid energy of $^{208}$Pb with that from Hartree-Fock + random phase
approximation calculations.

Furthermore, we have investigated how the ISGMR centroid energy difference
between $^{100}$Sn and $^{132}$Sn correlates with properties of asymmetric
nuclear matter. We have found that the ISGMR centroid energy difference
between $^{100}$Sn and $^{132}$Sn displays a strong correlation with the $L$
parameter while weak dependence on the other macroscopic quantities. This
feature implies that the ISGMR centroid energy difference between $^{100}$Sn
and $^{132}$Sn provides a potentially useful probe of the nuclear symmetry
energy. Our results also provide strong motivation for measuring the ISGMR
strength in unstable nuclei, which can be investigated at the new/planing
rare isotope beam facilities around the world.

\section*{ACKNOWLEDGMENTS}

This work was supported in part by the NNSF of China under Grant
Nos. 10975097, 10975190 and 11135011, Shanghai Rising-Star Program
under Grant No. 11QH1401100, ``Shu Guang" project supported by
Shanghai Municipal Education Commission and Shanghai Education
Development Foundation, the National Basic Research Program of China
(973 Program) under Contract Nos. 2007CB815003 and 2007CB815004, and
the Funds for Creative Research Groups of China under Grant No.
11021504.

\end{document}